\documentstyle[11pt]{article}
\def\fnote#1#2{\begingroup\def\thefootnote{#1}\footnote{#2}\addtocounter
{footnote}{-1}\endgroup}

\begin{document}

\hfill{UTTG-11-09}

\vspace{36pt}

\begin{center}
{\large {\bf {Asymptotically Safe Inflation}}}

\vspace{36pt}
Steven Weinberg\fnote{*}{Electronic address:
weinberg@physics.utexas.edu}\\
{\em Theory Group, Department of Physics, University of
Texas\\
Austin, TX, 78712}

\vspace{30pt}

\noindent
{\bf Abstract}
\end{center}
\noindent
Inflation is studied in the context of asymptotically safe theories of gravitation.  Conditions are explored under which it is possible  to have a long period of nearly exponential expansion that eventually comes to an end.
\vfill

\pagebreak

\begin{center}
{\bf I. Introduction}
\end{center}

Decades ago it was suggested that the effective quantum field theory of gravitation and matter might be asymptotically safe,\footnote{S. Weinberg,  in {\em Understanding the Fundamental Constituents of Matter}, ed. A. Zichichi (Plenum Press, New York, 1977).}  and hence ultraviolet-complete.  That is,  the renormalization group flows might have a fixed point, with a finite dimensional ultraviolet critical surface of trajectories attracted to the fixed point at short distances.  Evidence for a fixed point in the quantum theory of gravitation with or without matter has gradually accumulated through the use of dimensional continuation,\footnote{S. Weinberg, in {\em General Relativity}, ed. S. W. Hawking and W. Israel (Cambridge University Press, 1979): 700; H. Kawai, Y. Kitazawa, \& M. Ninomiya, Nucl. Phys. B 404, 684 (1993);   Nucl. Phys. B 467, 313 (1996); T.  Aida \& Y. Kitazawa, Nucl. Phys. B 401, 427 (1997);  M. Niedermaier, Nucl. Phys. B 673, 131 (2003) .}
 the large $N$ approximation\footnote{L. Smolin, Nucl. Phys. B208, 439 (1982);
 R. Percacci, Phys. Rev. D 73, 041501 (2006).} (where $N$ is the number of matter fields), lattice methods,\footnote{J. Ambj\o rn, J. Jurkewicz, \& R. Loll, Phys. Rev. Lett. 93, 131301 (2004);  Phys. Rev. Lett. 95, 171301 (2005); Phys. Rev. D72, 064014 (2005);   Phys. Rev. D78, 063544 (2008); and  in {\em Approaches to Quantum Gravity}, ed. D. Or\'{i}ti (Cambridge University Press).}  the truncated exact renormalization group,\footnote{M. Reuter, Phys. Rev. D 57, 971 (1998); M. Reuter, hep-th/9605030; D. Dou \& R. Percacci, Class. Quant. Grav. 15, 3449 (1998); W. Souma, Prog. Theor. Phys. 102, 181 (1999); O. Lauscher \& M. Reuter, Phys. Rev. D 65, 025013 (2001); Class. Quant. Grav. 19. 483 (2002);  M. Reuter \& F. Saueressig, Phys Rev. D 65, 065016  (2002); O. Lauscher \& M. Reuter, Int. J. Mod. Phys. A 17, 993 (2002);  Phys. Rev. D 66, 025026 (2002); M. Reuter and F. Saueressig, Phys Rev. D 66, 125001 (2002); R. Percacci \& D. Perini, Phys. Rev. D 67, 081503 (2002);  Phys. Rev. D 68, 044018 (2003); D. Perini, Nucl. Phys. Proc. Suppl. C 127, 185 (2004); D. F. Litim, Phys. Rev. Lett. {\bf 92}, 201301 (2004); A. Codello \& R. Percacci, Phys. Rev. Lett. 97, 221301 (2006); A. Codello, R. Percacci, \& C. Rahmede, Int. J. Mod. Phys. A23, 143 (2008);  M. Reuter \& F. Saueressig, 0708.1317; P. F. Machado and F. Saueressig, Phys. Rev. D77, 124045 (2008); A. Codello, R. Percacci, \& C. Rahmede, Ann. Phys. 324, 414 (2009);  A. Codello \& R. Percacci, 0810.0715; D. F. Litim 0810.3675; H. Gies \& M. M. Scherer, 0901.2459; D. Benedetti, P. F. Machado, \& F. Saueressig, 0901.2984, 0902.4630; M. Reuter \& H. Weyer, 0903.2971.  
For a review, see M. Reuter and P. Saueressig, to be published [0708.1317].} and a version  of perturbation theory.\footnote{M. R. Niedermaier, Phys. Rev. Lett. 103, 101303 (2009).} Recently there has also been evidence that the ultraviolet  critical surface is finite-dimensional; it has been found that even in truncations of the exact renormalization group equations with more than three (and up to nine) independent coupling parameters, the ultraviolet critical surface is just three-dimensional.\footnote{A. Codello, R. Percacci, \& C. Rahmede, Int. J. Mod. Phys. A23, 143 (2008); Ann. Phys. 324, 414 (2009); D. Benedetti, P. F. Machado, \& F. Saueressig, 0901.2984, 0902.4630}  The condition that physical parameters lie on the ultraviolet critical surface is analogous to the condition of renormalizability in the Standard Model, and like that condition yields a theory with a finite number of free parameters.

The natural arena for applications of the idea of asymptotic safety is the physics of very short distances, and in particular the early universe.\footnote{The implications of asymptotic safety for cosmology have been considered by A. Bonanno and M. Reuter, Phys. Rev. D 65, 043508 (2002); Phys. Lett. B527, 9 (2002); M.  Reuter and F. Saueressig, J. Cosm. and Astropart. Phys. 09, 012 (2005).  This work differs from that presented here, in that they consider a severe truncation of the gravitational action, including only the cosmological constant and Einstein--Hilbert terms; they include matter as a perfect fluid with a constant equation of state parameter $w$; and they employ a time-dependent cutoff $\Lambda$.  For more recent similar work that is  somewhat closer in spirit to the present paper, see A. Bonanno and M. Reuter, J. Cosm. and Astropart. Phys. 0708, 024 (2007); J. Phys. Conf. Ser. 140, 012008 (2008).}  In Section II we show how to formulate the differential equations for the scale factor in a Robertson--Walker solution of the classical field equations in a completely general generally covariant theory of gravitation.  In Section III we apply this result to  calculate the expansion rate $\overline{H}$ for a de Sitter solution of the classical field equations.    We are interested here in solutions for which $\overline{H}$ is of the same order as the scale at which the couplings are  beginning to approach their fixed point, or larger.  In this case,   $\overline{H}$ turns out in the tree approximation to depend strongly on the ultraviolet cutoff, indicating a breakdown of the classical approximation.  We deal with this by choosing an optimal cutoff, which minimizes the quantum corrections to the classical field equations.  Section IV considers more general time-dependent Robertson--Walker solutions of the classical field equations with an optimal cutoff, and explores the circumstances under which  it is possible to have an exponential expansion that persists for a long time but eventually comes to an end.  
An illustrative example is worked out in Section V.

We will work with a completely general generally covariant theory of gravitation.  (For simplicity matter will be ignored here.)   The effective action with an ultraviolet cutoff $\Lambda$ takes the form\footnote{Higher derivative theories of this sort if used in the tree approximation have long been known to be plagued by ``ghosts''; that is, poles in propagators with residues of the wrong sign for unitarity.  This is only if the series of operators in (1) is truncated; otherwise propagator denominators are not polynomials in the squared momentum, and there may be just one pole, or any number of poles.  Even with a truncated action,  because of the running of the couplings, there is no one Lagrangian that can be used to find the propagator in the tree approximation over the whole range of momenta where the various poles occur, and it is not ruled out that all the poles have the residues of the right sign.  For instance, ref. 6 shows that, in a theory with only the couplings  $g_1$, $g_{2a}$, and $g_{2b}$, the residue of the pole in the spin 2 propagator at high mass, which had usually been  supposed to have the wrong sign (as for instance in the work of K. S. Stelle, Phys. Rev. D 16, 953 (1977)), in fact has a sign consistent with unitarity.  More generally, Benedetti {\em et al.} in ref. 7  point out that for any truncation or no truncation, when we look for a pole at a four-momentum $p$, we must take the cut-off $\Lambda$ to be proportional to $\sqrt{-p^2}$, so the denominator of any propagator takes the form $p^2+m^2(-p^2)$.  The function $m^2(-p^2)$ is a constant at sufficiently low $|p^2|$, and of the form $cp^2$ for momenta so large that the couplings are near their fixed point, where $c$ is a constant, so the equation $p^2+m^2(-p^2)=0$ for the pole position has no solution if $-c>1$.}
\begin{eqnarray}
I_\Lambda [g] & =& -\int d^4x \,\sqrt{{-\rm Det} g}\Bigg[\Lambda^4 g_0(\Lambda)+\Lambda^2 g_1(\Lambda)R+g_{2a}(\Lambda)R^2\nonumber\\&& 
+g_{2b}(\Lambda)R^{\mu\nu}R_{\mu\nu}+\Lambda^{-2}g_{3a}(\Lambda)R^3+\Lambda^{-2}g_{3b}(\Lambda)RR^{\mu\nu}R_{\mu\nu} +\dots\Bigg]\;.
\end{eqnarray}
Here we have extracted powers of $\Lambda$ from the conventional coupling constants, to make the coupling parameters $g_n(\Lambda)$  dimensionless.  Because they are dimensionless, these running couplings satisfy renormalization group equations of the form
\begin{equation}
\Lambda\frac{d}{d\Lambda}g_n(\Lambda)=\beta_n \Big(g(\Lambda)\Big)\;.
\end{equation}
The condition for a fixed point at $g_n=g_{n*}$ is that $\beta_n(g_*)=0$ for all $n$.  As is well known, the condition for the couplings to be attracted to a fixed point $g_{n*}$ as $\Lambda\rightarrow\infty$ can be seen by considering the behavior of  $g_n(\Lambda)$ when it is near $g_{n*}$.  In the case where $\beta_n(g)$ is analytic in a neighborhood of  $g_{n*}$, near this fixed point we have
\begin{equation}
\beta_n(g)\rightarrow \sum_m B_{nm}\,\left(g_m-g_{*m}\right)\,~~~~~~B_{nm}\equiv \left(\frac{\partial\beta_n(g)}{\partial g_m}\right)_*\;,
\end{equation}
The solution of Eq.~(2) in this neighborhood is
\begin{equation}
g_n(\Lambda)\rightarrow g_{*n} +\sum_N u^N_n \left(\frac{\Lambda}{M}\right)^{\lambda_N}
\end{equation}
where $u^N$ and $\lambda_N$ are eigenvectors and corresponding eigenvalues of the matrix $B_{nm}$:
\begin{equation}
\sum_m B_{nm}\,u_m^N=\lambda_N\,u^N_n\;.
\end{equation}
It is a physical requirement that the only eigenvectors that are allowed to appear in the sum in Eq.~(4) are those for which the real part of the corresponding eigenvalues are negative, so that the couplings actually do approach the fixed point.  The normalizations of the eigenvectors that do appear in Eq.~(4) are free physical parameters, the only free parameters of the theory, except that we can adjust the over-all normalization of all the eigenvectors as we like by a suitable choice of the arbitrary mass scale $M$.  If we choose $M$ to make the largest of the $u_n^N$ of order unity, then $M$ is the cut-off scale at which couplings are just beginning to approach their fixed point.

Aside from the illustrative example considered in Section V, we will not carry our discussion in this paper to the point of performing numerical calculations, which of course would require some truncation of the series of terms in the action (1).  Our  purpose here is to lay out the general outlines of such a calculation, for which purpose we do not need to adopt any specific truncation.  Our results are worked out in detail for the terms explicitly shown in Eq.~(1), but this is only for the purposes of illustration; nothing in this paper assumes the neglect of higher terms.  For our purposes here, it makes no difference whether $\Lambda$ is regarded as a sharp ultraviolet cutoff on loop diagrams to be calculated using the action (1), or as a momentum parameter (usually called $k$) in a regulator term added to the action, or a sliding renormalization scale.

\begin{center}
{\bf II.  Robertson--Walker Solutions}
\end{center}

In this section we consider how to find a solution of the classical gravitational field equations for the general action (1), of the flat-space Robertson--Walker form
\begin{equation}
d\tau^2=dt^2-a^2(t)\,d\vec{x}^2\;.
\end{equation}
It would be very complicated to derive the ten classical  field equations for a general metric that follow from an action like (1), and then specialize to the case of a Robertson--Walker metric.  Instead, we can much more easily exploit the symmetries of this metric to derive a {\em single} differential equation for the Hubble rate $H(t)\equiv \dot{a}(t)/a(t)$.  In showing how to derive this differential equation, we will be  quite general, not making any use  in this section of the assumption of asymptotic safety.

We can use the rotational and translational symmetries of the line element (6) to write the components of the variational derivatives $\delta I_\Lambda/\delta g_{\mu\nu}$ in the form
\begin{eqnarray}
&&\left[\frac{\delta I_\Lambda[g]}{\delta g_{ij}(x)}\right]_{\rm RW}=\frac{\Lambda^4}{6}\delta_{ij}\, a^{-2}(t)\,{\cal M}_\Lambda(t)\;,\\
&&\left[\frac{\delta I_\Lambda[g]}{\delta g_{i0}(x)}\right]_{\rm RW}=0\\
&&\left[\frac{\delta I_\Lambda[g]}{\delta g_{00}(x)}\right]_{\rm RW}=-\frac{\Lambda^4}{2}{\cal N}_\Lambda(t)\;,
\end{eqnarray}
the subscript RW indicating that, after taking the variational derivative, the metric is to be set 
 equal to the 
Robertson--Walker metric defined by (6).  (The factors $\Lambda^4/6a^2$ and $\Lambda^4/2$ are inserted in the definitions of ${\cal M}_\Lambda$ and ${\cal N}_\Lambda$  for future convenience.) 
Also, the general covariance of the action yields the generalized Bianchi identity
\begin{equation}
0=\left[\frac{\delta I_\Lambda[g]}{\delta g_{\mu\nu}(x)}\right]_{;\nu}\;.
\end{equation}
By using  Eqs.~(7)--(9) for the Robertson--Walker metric, Eq.~(10) is reduced to the condition:
\begin{equation}
a^2\dot{a}\,{\cal M}_\Lambda=\frac{d}{d t}\Big(a^3\,{\cal N}_\Lambda\Big)\;.
\end{equation}
Therefore the gravitational field equations reduce here to a single  differential equation:
\begin{equation}
{\cal N}_\Lambda(t)=0\;,
\end{equation}
which we see ensures the vanishing of all variational derivatives $\delta I_\Lambda[g]/\delta g_{\mu\nu}$.
This result (which holds also in the presence of spatial curvature and matter) is the generalization of the familiar Friedmann equation,  which would apply if only the Einstein--Hilbert term $-\sqrt{g}R/16\pi G$ and a vacuum energy term were included in the gravitational action.

We can express ${\cal M}_\Lambda$ and then ${\cal N}_\Lambda$ in terms of variational derivatives of the action for the Robertson-Walker metric with respect to the scale factor $a(t)$.  Because $a(t)$ appears in the Robertson--Walker metric only as a factor $a^2(t)$ in $g_{ij}({\bf x},t)$, we have
\begin{equation}
\frac{\delta I_\Lambda[g_{{\rm RW}}]}{\delta a(t)}=\int d^3x\; 2a(t)\delta_{ij}\times a^3(t)\,\left[\frac{\delta I[g]}{\delta g_{ij}({\bf x},t)}\right]_{\rm RW}=V\,\Lambda^4\,{\cal M}_\Lambda(t)\,a^2(t)\;,
\end{equation}
where $V$ is the coordinate space volume (which can be made finite by imposing periodic boundary conditions.) For the flat-space Robertson-Walker metric $(g_{\rm RW})_{\mu\nu}$, the action takes the general form
\begin{equation}
I_\Lambda[g_{{\rm RW}}]=V\Lambda^4\int dt\;a^3(t)\,{\cal I}_\Lambda\Big(H(t),\dot{H}(t),\dots\Big)\;,
\end{equation}  
where as usual $H(t)\equiv \dot{a}(t)/a(t)$.
Here and in Eqs.~(15)--(17) below,  the ellipsis $\dots$  indicates a possible dependence of ${\cal I}_\Lambda$ on second and higher derivatives of $H(t)$.  (Second and higher time derivatives  do not occur in ${\cal I}_\Lambda$ if the integrand of the action is $\sqrt{-{\rm Det}\,g}$ times an arbitrary scalar function of the Riemann-Christoffel curvature tensor $R_{\mu\nu\rho\sigma}$, including of course an arbitrary dependence on the curvature scalar and the Ricci tensor, but we do not assume that this is the case.)
Comparing Eq.~(13) with the result of a straightforward calculation of the variational derivative of the action (14) with respect to $a(t)$  gives
\begin{eqnarray}
{\cal M}_\Lambda&= & 3{\cal I}_\Lambda-3H\frac{\partial{\cal I}_\Lambda}{\partial H}+(3\dot{H}+9H^2)\frac{\partial{\cal I}_\Lambda}{\partial \dot{H}}\nonumber\\&&-\frac{d}{dt}\left(\frac{\partial {\cal I}_\Lambda}{\partial H}\right)
+6H\frac{d}{dt}\left(\frac{\partial {\cal I}_\Lambda}{\partial \dot{H}}\right)+\frac{d^2}{dt^2}\left(\frac{\partial {\cal I}_\Lambda}{\partial \dot{H}}\right)+\dots\;.~~~
\end{eqnarray}
We note that $a^2\dot{a}{\cal M}_\Lambda$ is a time-derivative
\begin{eqnarray}
a^2\dot{a}{\cal M}_\Lambda&= & \frac{d}{dt}\Bigg\{a^3\Bigg[{\cal I}_\Lambda-H\frac{\partial{\cal I}_\Lambda}{\partial H}+(-\dot{H}+3H^2)\frac{\partial{\cal I}_\Lambda}{\partial \dot{H}}\nonumber\\&&+H\frac{d}{dt}\left(\frac{\partial {\cal I}_\Lambda}{\partial \dot{H}}\right)+\dots\Bigg]\Bigg\}
\;.~~~
\end{eqnarray}
Comparing with Eq.~(11), we see that ${\cal N}_\Lambda$ equals the term in square brackets in (16), up to a possible term equal to a constant divided by $a^3(t)$.  But the term in square brackets is independent of the scale of $a(t)$, as is ${\cal N}_\Lambda(t)$, so there can be no term in their difference proportional to $1/a^3(t)$, and thus
\begin{equation}
{\cal N}_\Lambda={\cal I}_\Lambda-H\frac{\partial{\cal I}_\Lambda}{\partial H}+(-\dot{H}+3H^2)\frac{\partial{\cal I}_\Lambda}{\partial \dot{H}}+H\frac{d}{dt}\left(\frac{\partial {\cal I}_\Lambda}{\partial \dot{H}}\right)+\dots
\end{equation}
The ten classical field equations reduce for the flat-space Robertson--Walker metric to the single requirement that this vanishes.

To evaluate the terms in the action for the Robertson--Walker metric with no spatial curvature that are explicitly shown in Eq.~(1),  we note that for this metric $R=-12H^2-6\dot{H}$ and $R_{\mu\nu}R^{\mu\nu}=36H^4+36H^2\dot{H}+12\dot{H}^2$.  Using these in Eq.~(1) and comparing with Eq.~(14) gives
\begin{eqnarray}
&&{\cal I}_\Lambda=-g_0(\Lambda)+\Lambda^{-2} g_1(\Lambda)(12H^2+6\dot{H})-\Lambda^{-4}g_{2a}(\Lambda)(12H^2+6\dot{H})^2\nonumber\\&&~~~
-\Lambda^{-4}g_{2b}(\Lambda)\left(36H^4+36H^2\dot{H}+12\dot{H}^2\right)+\Lambda^{-6}g_{3a}(\Lambda)(12H^2+6\dot{H})^3\nonumber\\&&~~~+\Lambda^{-6}g_{3b}(\Lambda)(12H^2+6\dot{H})(36H^4+36H^2\dot{H}+12\dot{H}^2)+\dots\;,~~~~
\end{eqnarray}
where now the dots  $\dots$ denote contributions from terms not shown in (1), some of which involve second and higher derivatives of $H$.  From Eq.~(17), we then have 
\begin{eqnarray}
&&{\cal N}_\Lambda(H,\dot{H},\ddot{H},\dots)=-g_0(\Lambda)+6\Lambda^{-2}g_1(\Lambda)H^2\nonumber\\&&~~-\Lambda^{-4}g_{2a}(\Lambda)\Big(216H^2\dot{H}-36\dot{H}^2+72H\ddot{H}\Big)\nonumber\\&&~~~-\Lambda^{-4}g_{2b}(\Lambda)\Big(72H^2\dot{H}-12\dot{H}^2+24H\ddot{H}\Big)\nonumber
\\&&~~~+\Lambda^{-6}g_{3a}(\Lambda)\Big(-864\,H^6+7776H^4\dot{H}+3240H^2\dot{H}^2\nonumber\\&&~~~~~~~~~-432\,\dot{H}^3+216\,H\ddot{H}(12H^2+6\dot{H})\Big)\nonumber\\&&~~~+\Lambda^{-6}g_{3b}(\Lambda)\Big(-216H^6+2160H^4\dot{H}
+1008\,H^2\dot{H}^2-144\dot{H}^3\nonumber\\&&~~~~~+H\ddot{H}(720H^2+432\dot{H})\Big)+\dots\;.
\end{eqnarray}
This is the quantity that must be set equal to zero in finding a flat-space Robertson--Walker solution of the classical gravitational field equations.

\begin{center}
{\bf III.  De Sitter Solutions and Optimal Cutoff}
\end{center}

We can now easily find the condition for a de Sitter solution of the classical field equations, with 
\begin{equation}
a(t)\propto e^{\overline{H}t}\;,
\end{equation}
where $\overline{H}$ is constant.  Setting the quantity (19) equal to zero  for $H(t)=\overline{H}$  gives our condition on $\overline{H}$:\footnote{Note that this is not the result that would be obtained by setting the derivative of  ${\cal I}_\Lambda(\overline{H},0,0,\dots)$ with respect to $\overline{H}$ equal to zero.  For a de Sitter metric with $a(t)=\exp(\overline{H}t)$, the integral over $t$ in the action $I_\Lambda[g]$ diverges at $t=\infty$.  If we integrate only from $t=-\infty$ to $t=0$, the integral $\int dt\,a^3(t)$ gives a factor $1/3\overline{H}$, but the derivative of $ {\cal I}_\Lambda(\overline{H},0,0,\dots)/3\overline{H}$ with respect to $\overline{H}$ is not zero; it equals a surface term $(\partial {\cal I}_\Lambda/\partial \dot{H})_{\overline{H}}$, which again gives Eq.~(21).}  
\begin{eqnarray}
0&=&N_\Lambda(\overline{H})\equiv {\cal N}_\Lambda(\overline{H},0,0,\dots)\nonumber\\&=&-g_0(\Lambda)+6\,g_1(\Lambda)\,(\overline{H}/\Lambda)^2
-864\,g_{3a}(\Lambda)\,(\overline{H}/\Lambda)^6\nonumber\\&&-216\,g_{3b}(\Lambda)\,(\overline{H}/\Lambda)^6+\dots
\end{eqnarray}

It is easy to find solutions of Eq.~(21) that have  small values of $\overline{H}$, very much smaller than the scale $M$ at which the couplings begin to approach their fixed points.  For sufficiently small $\overline{H}$, we can take $\Lambda$ to be much larger than $\overline{H}$, and yet small enough so that the couplings  appearing as coefficients in (1) become independent of $\Lambda$, and in particular
$$
\Lambda^4 g_0(\Lambda)\rightarrow \rho_V\;,~~~\Lambda^2 g_1(\Lambda)\rightarrow 1/16\pi G_N\;,
$$
where $\rho_V$ and $G_N$ are the conventional, $\Lambda$-independent, vacuum energy and Newton constant.  Then (21) has the familiar $\Lambda$-independent solution
$$
\overline{H}^2=\frac{8\pi G_N\rho_V}{3}\;.
$$
Because of the still mysterious fact that $\rho_V$ is observed to be much less than $G^{-2}$, this value of $\overline{H}$ is much less than $G^{-1/2}$, and so radiative corrections and higher terms in (21) can be neglected.

We will instead be interested here in looking for solutions for which $\overline{H}$ is roughly of the order of the 
scale $M$ at which the couplings begin to approach their fixed points, or larger.  In this case, we face a difficult choice:   How should we choose $\Lambda$?  
On one hand, if we choose $\Lambda\ll \overline{H}$, then we can expect radiative corrections to the classical result (21) to be unimportant, because $\overline{H}$ provides a natural infrared cutoff in loop diagrams constructed using the action (1).  But for $\Lambda\ll \overline{H}$, the sum (21) receives increasing contributions as we include   higher and higher terms, and whether or not the series actually converges, it is not useful.  On the other hand, if we choose $\Lambda\gg \overline{H}$, then it is reasonable to suppose that the series (21) is dominated by its lowest terms, but for $\Lambda\gg \overline{H}$ there is no reason to suppose that we can neglect radiative corrections to the field equations.  Indeed, we can see that radiative corrections to the field equations {\em are} important here, because where Eq.~(21) is dominated by its lowest terms, it gives $\overline{H}$ a strong dependence on $\Lambda$.  (This is clearest in the case where $\Lambda$ is so large that the couplings are near their fixed points, in which case (21) gives $\overline{H}$ proportional to $\Lambda$.) The whole point of the renormalization group equations (2) is that physical quantities like $\overline{H}$ should be independent of the cutoff, but in general this is true only when radiative corrections are included, and since Eq.~(21) gives  $\overline{H}$ a strong dependence on $\Lambda$ when $\Lambda\gg \overline{H}$, radiative corrections evidently can not be neglected.  

Ideally, we should leave $\Lambda$ undetermined, and calculate enough of the  radiative corrections to the field equations  so that $\overline{H}$ comes out at least approximately independent of $\Lambda$.  This would not be easy.  Instead, we can try to make a judicious choice of  $\Lambda$ to minimize the radiative corrections.  We can guess that the optimal $\Lambda$  is roughly of the order of $\overline{H}$, where radiative corrections are just beginning to be important, and the higher terms in (21) are just beginning to be less important.  This sort of guess works quite well in quantum chromodynamics.  The radiative corrections to a process like $e^+$--$e^-$ annihilation into jets of hadrons at an energy $E$ are accompanied with powers of $\ln(E/\Lambda)$, and to avoid large radiative corrections it is only necessary to take $\Lambda\approx E$.  In this way, we can use the tree approximation to calculate the annihilation into, say, three jets, with the renormalization scale of the QCD coupling taken of order $E$.  But in our case, radiative corrections are more sensitive to $ \Lambda$, and we have to make a more careful choice of $\Lambda$.

To find an optimal cutoff, we note that in principle we should  find $\overline{H}$ by solving the full  quantum corrected field equations, which give a result that can be schematically written as 
\begin{equation}
\overline{H}_{\rm true}= \overline{H}(\Lambda)+\Delta \overline{H}(\Lambda)\;,
\end{equation}
where $ \overline{H}(\Lambda)$ is defined as the solution of Eq.~(21), and $ \Delta \overline{H}(\Lambda)$ represents the effect of radiative corrections.   
Instead of calculating loop graphs,  we can get some idea of the results of such a calculation  by using the tree-approximation field equations (21), but with $\Lambda$ chosen at a local  minimum of the radiative corrections to  $\overline{H}$.   For such an optimal  $\Lambda$, we have\footnote{This is the weakest point in our discussion.  For one thing, we do not know whether the condition (23) gives a local minimum or maximum of the radiative corrections.  Worse, even if the radiative corrections are minimized, we do not know that they are small.} 
\begin{equation}
\frac{\partial}{\partial \Lambda} \Delta \overline{H}(\Lambda)=0\;.
\end{equation}
As already mentioned, physical quantities, including the  true expansion rate $\overline{H}_{\rm true}$,   must be independent of $\Lambda$, so Eq.~(23) tells us also that the expansion rate calculated from the classical field equations is  stationary at the optimal cut-off
\begin{equation}
0= \Lambda \frac{\partial}{\partial \Lambda} \overline{H}(\Lambda)\;.
\end{equation}
By definition, for any $\Lambda$ we have $N_\Lambda\Big(\overline{H}(\Lambda)\Big)=0$, and by differentiating this with respect to $\Lambda$ and using Eq.~(24) we find that the condition for an optimal cutoff may be put in the form
\begin{equation}
0=\left.\Lambda\frac{\partial}{\partial \Lambda}N_\Lambda(\overline{H})\right|_{\overline{H}=\overline{H}(\Lambda)}
=A_\Lambda\Big(\overline{H}(\Lambda)\Big)+B_\Lambda\Big((\overline{H}(\Lambda)\Big)\;,
\end{equation}
where $A_\Lambda$ arises from the explicit  dependence of $ N_\Lambda(\overline{H})$ on $\overline{H}/\Lambda$:
\begin{eqnarray}
A_\Lambda(\overline{H})&\equiv & -\overline{H}\frac{\partial}{\partial \overline{H}} N_\Lambda(\overline{H}) \nonumber \\&&=-12\left(\frac{\overline{H}}{\Lambda}\right)^{2}g_1(\Lambda)+5184\left(\frac{\overline{H}}{\Lambda}\right)^{6}g_{3a}(\Lambda)\nonumber\\&& ~~~~~~~~+1296\,\left(\frac{\overline{H}}{\Lambda}\right)^{6}g_{3b}(\Lambda) +\dots\;,
\end{eqnarray}
and $B_\Lambda$ comes from the running of the couplings in $N_\Lambda$:
\begin{eqnarray}
B_\Lambda(\overline{H})&\equiv& -\beta_0\Big(g(\Lambda)\Big)+6\,\beta_1\Big(g(\Lambda)\Big)\,(\overline{H}/\Lambda)^2
-864\,\beta_{3a}\Big(g(\Lambda)\Big)\,(\overline{H}/\Lambda)^6\nonumber\\&& -216\,\beta_{3b}\Big(g(\Lambda)\Big)\,(\overline{H}/\Lambda)^6 +\dots\;.~~~~~~~
\end{eqnarray}
We now have two equations, (21) and (25), for the two  quantities $\overline{H}$ and $\Lambda$, so it is not unreasonable to expect there to be one or more solutions, with both $\Lambda$ and $\overline{H}$ roughly of order $M$, the only mass parameter in the theory.

\begin{center}
{\bf IV. Time Dependence}
\end{center}

The de Sitter solution found in Section II describes a universe that inflates eternally.  For a more realistic picture of inflation, we need a solution that remains close to the de Sitter solution with expansion rate near $\overline{H}$ for a time much longer than $1/\overline{H}$, but that gradually evolves away from the de Sitter solution, so that inflation can come to an end.  (We have nothing to  say here about the metric before the  universe enters into its de Sitter phase.)  To find such a solution, we will consider first-order perturbations of the de Sitter solution, of the Robertson--Walker form (6).  The expansion rate will take the form
\begin{equation}
H(t)=\overline{H}+\delta H(t)\;,
\end{equation}
with $|\delta H(t)|\ll \overline{H}$.  Keeping only terms in (19) of first order in $\delta H(t)$, the field equation  ${\cal N}_\Lambda=0$ becomes
\begin{equation}
c_0(\overline{H},\Lambda)\frac{\delta H}{\overline{H}} +c_1(\overline{H},\Lambda)\frac{\delta\dot{H}}{\overline{H}^2}+c_2(\overline{H},\Lambda)\frac{\delta\ddot{H}}{\overline{H}^3}+\dots=0\;,
\end{equation}
where
\begin{equation}
c_0(\overline{H},\Lambda)\equiv\overline{H}\left(\frac{\partial{\cal N}_\Lambda}{\partial H}\right)_{\overline H}=-A_\Lambda(\overline{H})\;,
\end{equation}
with $A_\Lambda$ given by Eq.~(26), and
\begin{eqnarray}
&&c_1(\overline{H},\Lambda)\equiv\overline{H}^2\left(\frac{\partial{\cal N}_\Lambda}{\partial\dot{H}}\right)_{\overline H}\nonumber\\&&~~~=-216\, g_{2a}(\Lambda)\left(\frac{\overline{H}}{\Lambda}\right)^4-72\, g_{2b}(\Lambda)\left(\frac{\overline{H}}{\Lambda}\right)^4+7776\,g_{3a}(\Lambda)\left(\frac{\overline{H}}{\Lambda}\right)^6\nonumber\\&&~~~~~~~~~+2160\,g_{3b}(\Lambda)\left(\frac{\overline{H}}{\Lambda}\right)^6+\dots \;,
\end{eqnarray}
\begin{eqnarray}
&&c_2(\overline{H},\Lambda)\equiv\overline{H}^3\left(\frac{\partial{\cal N}_\Lambda}{\partial\ddot{H}}\right)_{\overline {H}}\nonumber\\&&~~~=-72 \,g_{2a}(\Lambda)\left(\frac{\overline{H}}{\Lambda}\right)^4-24\, g_{2b}(\Lambda)\left(\frac{\overline{H}}{\Lambda}\right)^4+2592\,g_{3a}(\Lambda)\left(\frac{\overline{H}}{\Lambda}\right)^6\nonumber\\&&~~~~~~~~~+720\,g_{3b}(\Lambda)\left(\frac{\overline{H}}{\Lambda}\right)^6+\dots \;,
\end{eqnarray}
and so on, with the subscript $\overline{H}$ on partial derivatives meaning that after taking the derivatives we set $H(t)=\overline{H}$.  Eq.~(29) has an obvious solution of the form
\begin{equation}
\delta H\propto \exp(\xi\overline{H} t)\;,
\end{equation}
where $\xi$ is any root of the equation
\begin{equation}
c_0(\overline{H},\Lambda) +c_1(\overline{H},\Lambda)\,\xi+c_2(\overline{H},\Lambda)\,\xi^2+\dots=0\;.
\end{equation}
(This is a quadratic equation in the special case in which the integrand of the action is $\sqrt{-{\rm Det}g}$ times an arbitrary function of the curvature tensor.)
For positive ${\rm Re}\;\xi$, Eq.~(33) represents an instability, and the number of $e$-foldings before this instability ends the exponential expansion is $\approx 1/{\rm Re}\;\xi$.

We would generally expect the  coefficients in Eq.~(34)  to be of the same order, in which case typical solutions for $\xi$ would  be of order unity, and  inflation would either end almost immediately (if ${\rm Re}\;\xi>0$) or go on forever (if ${\rm Re}\;\xi\leq 0$).  But there are various circumstances under which we expect $\xi$ to be much smaller, giving a large number of $e$-foldings before the end of inflation.\footnote{We are concentrating here on only one mode.  In all cases Eq.~(34) will have more than one solution, and we are assuming that all modes other than the one (or several) with
${\rm Re}\;\xi$ small and positive either have ${\rm Re}\;\xi\leq 0$ or for some reason are not excited.} 
\begin{enumerate}
\item If $|c_0|$ is much less than all the other $|c_n|$, then  Eq.~(34) will  have a solution 
$\xi\simeq -c_0/c_1$,  and so much less than unity.  In particular, if we now choose $\Lambda$ to be the optimal cutoff described in the previous section, then we can use the condition (25) and Eq.~(30) to write 
\begin{equation}
c_0(\overline{H},\Lambda)=B_\Lambda(\overline{H})\;,
\end{equation}
According to Eq.~(27), $B_\Lambda(H)$ vanishes if the couplings are at their  fixed point, so we can conclude that it  is possible to have  a long but not eternal period of inflation if the optimal $\Lambda$ is large enough so that the couplings $g_n(\Lambda)$ are not far from  their fixed point.  But there is a limit to how close the couplings at the optimum cutoff can be to their fixed point.  At the fixed point, the quantities (21) and (25) are both functions of the single parameter $\overline{H}/\Lambda$, and it is not likely that these two functions would vanish at the same value of this parameter.  
\item If the couplings are not very near their fixed point, they are sensitive to the free parameters of the theory that characterize the particular trajectory in coupling-constant space on which the couplings lie, and it is easy to choose these couplings  to make $|c_0|$ as small as we like.  For instance, where (4) applies, all the couplings are linear in the normalization of the eigenvectors $u_n^N$, the only free parameters of the theory.  In a theory of chaotic inflation, the value of these parameters in any big bang containing observers may be conditioned by the requirement that $c_0$ should be small enough (and have the right sign) to allow the bang to become big.  To be specific, in order for spatial curvature not to interfere with the formation of galaxies it is necessary that the universe should expand enough during inflation so that whatever curvature was present at the beginning of inflation would be decreased enough so that the curvature term in the Friedmann equation should not dominate over the matter term when galaxies form.\footnote{B. Freivogel, M, Kleban, M. R. Martinez, and L. Susskind, J. High Energy Phys. 0603, 039 (2006).} As is well known, the fact that spatial curvature does not dominate at present requires about 60 
to 70 $e$-foldings of inflation,\footnote{A. Guth, Phys. Rev. D23, 347 (1981).} and the anthropic requirement that curvature does not interfere with galaxy formation is almost as restrictive.  But the combination of data from the microwave background, baryon acoustic oscillations, and type Ia supernovae distance--redshift relations has shown\footnote{E. Komatsu {\em et al.}, Astrophys. J. Suppl. Ser. 180, 330 (2009).} that (within two standard deviations) the fractional curvature contribution $\Omega_K$ to $H_0^2$ is in the range of $-0.0178$ to $+0.0066$.  It is hard to see any anthropic reason for a number of $e$-foldings large enough to reduce the curvature this much.
\item Instead of $c_0$ being anomalously small, it is possible for some or all of the other $c_n$ to be anomalously large, in which  case again $\xi$ will be small and the number of $e$-foldings will be large.  For instance, we note that $c_0$ unlike the other $c_n$ does not involve the couplings $g_{2a}$ and $g_{2b}$, so if these couplings are anomalously large, as in ref. 6, then $c_1$, $c_2$, etc., will be much larger than $c_0$, and again we will have $|\xi|\simeq |c_0/c_1|\ll 1$.
\end{enumerate}
   
\begin{center}
{\bf V. An Example}
\end{center}

We will now apply the above results to a classic example of higher derivative theories of gravitation, with action limited 
to terms with no more than four spacetime derivatives:
\begin{eqnarray}
I_\Lambda [g] & =& -\int d^4x \,\sqrt{{-\rm Det} g}\Bigg[\Lambda^4 g_0(\Lambda)+\Lambda^2 g_1(\Lambda)R+g_{2a}(\Lambda)R^2\nonumber\\&& 
+g_{2b}(\Lambda)R^{\mu\nu}R_{\mu\nu}\Bigg]\;.
\end{eqnarray}
This theory was studied by Stelle\footnote{K. S. Stelle, ref. 9.} as a possible renormalizable quantum theory of gravitation, and has been considered recently by Niedermaier\footnote{M. R. Niedermaier, ref. 6.} and by Benedetti {\em et al.}\footnote{D. Benedetti, P. F. Machado, and F. Saueressig, ref. 7.} in connection with asymptotic safety.  As is well known, it is possible by using the Gauss--Bonnet identity to put this action in the form used in refs. 17 and 18:
\begin{eqnarray}
I_\Lambda [g] & =& -\int d^4x \,\sqrt{{-\rm Det} g}\Bigg[\Lambda^4 g_0(\Lambda)+\Lambda^2 g_1(\Lambda)R+f_{a}(\Lambda)R^2\nonumber\\&& 
+f_{b}(\Lambda)C^{\mu\nu\rho\sigma}R_{\mu\nu\rho\sigma}\Bigg]\;,
\end{eqnarray}
where $C_{\mu\nu\rho\sigma}$ is the Weyl tensor, and 
\begin{equation}
f_a=g_{2a}+\frac{g_{2b}}{3}\,,~~~~~f_b=\frac{g_{2b}}{2}\;.
\end{equation}

For this action, Eq.~(21) gives the expansion rate for a de Sitter solution of the field equations as
\begin{equation}
\overline{H}=\Lambda\sqrt{g_0(\Lambda)/6g_1(\Lambda)}\;.
\end{equation}
Instead of trying to find an optimal value of $\Lambda$, which minimizes radiative corrections to Eq.~(39), here we will simply assume that $\Lambda$ is large enough so that the couplings $g_n(\Lambda)$ are near their fixed point $g_{n*}$, and use Eq.~(39) to express $\Lambda$ in terms of $\overline{H}$:
\begin{equation}
\Lambda=\overline{H}\sqrt{6g_{1*}/g_{0*}}\;,
\end{equation}
with $\overline{H}$ left undetermined.

The critical question for this sort of theory is whether the de Sitter solution has an instability that ends the eepxonential expansion after a finite but large number of $e$-foldings.  As we have seen, for any small perturbation of the de Sitter solution, $\dot{a}/a$ is a sum of terms with the time dependence $\exp(\xi\overline{H}t)$, with $\xi$ running over the roots of Eq.~(34).  We are now considering an action whose integrand is $\sqrt{-{\rm Det}g}$ times a scalar function of the metric and the Riemann--Christoffel curvature tensor, so as remarked in the previous section, this equation is quadratic:
\begin{equation}
c_0+c_1\xi+c_2\xi^2=0\,.
\end{equation}
For the particular action (36), the coefficients are given by
\begin{equation}
c_0=12\,g_{1*}\,(\overline{H}/\Lambda)^2=2g_{0*}
\end{equation}
\begin{equation}
c_1=3c_2=\Big(-216g_{2a*}-72g_{2b*}\Big)(\overline{H}/\Lambda)^4=\Big(-6g_{2a*}-2g_{2b*}\Big)g_{0*}^2/g_{1*}^2\;,
\end{equation}
so Eq.~(41) reads
\begin{equation}
\xi^2+3\xi=A\;,
\end{equation}
where 
\begin{equation}
A=-\frac{c_0}{c_2}=\frac{3g_{1*}^2}{g_{0*}\,(3\,g_{2a*}+g_{2b*})}\;.
\end{equation}
We get a realistic picture of inflation if it turns out that $A$ is small and positive.  In this case Eq.~(44) has a root with $\xi\simeq -3$, corresponding to a perturbation to $\dot{a}/a$ that decays as $\exp(-3\overline{H}t)$, and a root with $\xi\simeq A/3$, corresponding to a slowly growing perturbation, that ends the exponential phase after about $3/A$ 
$e$-foldings.

Unfortunately, the numerical results obtained in ref. 17 and 18 are not encouraging.  The calculations of ref. 17 are expressed in terms of coupling constants $\lambda$, $g_N$, $\omega$, and $s$, related to the couplings in Eq.~(36) by
\begin{eqnarray}
&& g_0=2\lambda/g_N\;,~~~~~g_1=1/g_N \nonumber\\
&& g_{2a}=-(1+\omega)/3s\;,~~~~~g_{2b}=1/s\;.
\end{eqnarray}
Using a version of perturbation theory, ref. 17 found that for $\Lambda\rightarrow \infty$ the parameters $\omega$, $\lambda$ and $g_N$ approach the fixed point values
\begin{equation}
\omega_*=-0.0228\,,~~~~\lambda_*=12.69\,~~~~g_{N*}/(4\pi)^2=0.4227\;,
\end{equation}
while $s(\Lambda)$ vanishes as 
\begin{equation}
s(\Lambda) \rightarrow 11.88/\ln(\Lambda/M)\;,
\end{equation}
where $M$ is some unknown large mass.  Then Eq.~(45) gives
\begin{equation}
A=-\frac{3s}{2\omega\lambda g_N}\rightarrow \frac{0.92}{\ln(\Lambda/M)}\;,
\end{equation}
so $A$ is positive, but $\Lambda/M$ would have to be about $10^8$ to give 60 $e$-foldings before inflation ends.

In ref. 18, by using the truncated exact renormalization group equations, a fixed point is found with (in our notation)
\begin{equation}
g_{0*}=-0.0042\,,~~g_{1*}=-0.0101\,,~~g_{2a*}=-0.0109\,,~~g_{2b*}=0.01\;.
\end{equation}
Using these results in Eq.~(45) gives $A=3.05$.  This is positive, but unfortunately not at all small.  The two roots of Eq.~(44) are $\xi=-3.80$, correspondign to a rapidly decaying mode, and $\xi=0.80$, corresponding to an instability that ends inflation after only a few $e$-foldings.

\vspace{15pt}

I am grateful for discussions with D. Benedetti, W. Fischler, E. Komatsu, M. Niedermaier, and M. Reuter.  This material is based in part on work supported by the National Science Foundation under Grant NO. PHY-0455649 and with support from The Robert A. Welch Foundation, Grant No. F-0014.

\end{document}